\documentclass{article}

\usepackage{arxiv}

\usepackage[utf8]{inputenc} 
\usepackage[T1]{fontenc}    
\usepackage{hyperref}       
\usepackage{url}            
\usepackage{booktabs}       
\usepackage{amsfonts}       
\usepackage{nicefrac}       
\usepackage{microtype}      
\usepackage{lipsum}		
\usepackage{amssymb,amsmath}
\usepackage{listings}

\title{Efficient algorithms for modifying and sampling from a categorical distribution}

\author{
  Daniel Tang\\
  Leeds Institute for Data Analytics\\
  University of Leeds\\
  Leeds, UK\\
  \texttt{D.Tang@leeds.ac.uk} \\
}

\begin{document}
\maketitle

\begin{abstract}
Probabilistic programming languages and other machine learning applications often require samples to be generated from a categorical distribution where the probability of each one of $n$ categories is specified as a parameter. If the parameters are hyper-parameters then they need to be modified, however, current implementations of categorical distributions take $\mathcal{O}(n)$ time to modify a parameter. If $n$ is large and the parameters are being frequently modified, this can become prohibitive. Here we present the insight that a Huffman tree is an efficient data structure for representing categorical distributions and present algorithms to generate samples as well as add, delete and modify categories in $\mathcal{O}(\log(n))$ time. We demonstrate that the time to sample from the distribution remains, in practice, within a few percent of the theoretical optimal value. The same algorithm may also be useful in the context of adaptive Huffman coding where computational efficiency is important.

\end{abstract}


\section{Introduction}

With the recent rise in popularity of probabilistic programming libraries such as PyMC3\cite{salvatier2016probabilistic} and Tensorflow Probability\cite{dillon2017tensorflow} there is a need to develop efficient algorithms to work with probability distributions. One type of probability distribution commonly used in probabilistic programming is the \textit{categorical distribution} (sometimes called an \textit{empirical distribution}) which is a probability distribution over a finite number of states where the probability of each state is specified as a parameter, i.e. a categorical distribution, $P_\mathbf{\pi}$, over $n$ states $1 \dots n$ with a parameter $\mathbf{\pi} = \left<\pi_1, \pi_2 \dots \pi_n \right>$ has a probability mass function
\[
P_\mathbf{\pi}(i) = \pi_i
\]
Sampling from this distribution is a fundamental operation required in probabilistic programming (e.g. for sample-based inference). The standard algorithm to do this\footnote{as used in, for example, in PyMC3 and NumPy} is to first calculate the cumulative mass function
\[
C_\mathbf{\pi}(i) = \sum_1^i \pi_i
\]
then generate a uniformly distributed random number, $x$, in the range $[0,1)$ and finally use binary search to find the smallest $i$ such that $x < C_\mathbf{\pi}(i)$. Although not optimal, this algorithm is appropriate as long as the parameter $\mathbf{\pi}$ is fixed. However, in the context of probabilistic programming, $\pi$ is often a hyper-parameter which may itself change between samples. In this case, the cumulative mass function needs to be re-calculated at an amortised computational cost of $\mathcal{O}(n)$. If $n$ is large, and $\pi$ changes frequently, this can be prohibitively expensive.

If all elements of $\pi$ change we cannot do better than $\mathcal{O}(n)$, however quite often $\pi$ is subject to small perturbations (for example, when performing particle filtering involving categorical distributions with hyper-parameters or using an adaptive proposal distribution in MCMC). Here we present an algorithm that allows $\pi$ to be changed by adding, removing or changing the probability of any parameter. Each of these operations is performed in $\mathcal{O}(\log(n))$ time. The cost of taking a sample has the same complexity as the standard algorithm, $\mathcal{O}(\log(n))$, although we show that the constant of proportionality of our algorithm is often smaller.

\section{Huffman coding trees}

A binary search tree is not the optimal binary tree to transform a uniformly distributed sample into a sample from a categorical distribution. To see this intuitively, suppose for example that we have 1024 categories where $\pi_{1\dots 1023} = \frac{0.1}{1023}$ and $\pi_{1024} = 0.9$. Consider now the average number of branches that need to be traversed in a single lookup. In a binary tree, since there are 1024 items, we always need to traverse 10 branches. However, consider now a tree where the first left branch from the root is a leaf-node representing the $1024^{th}$ category and the right branch from the root leads on to a binary search tree of the remaining 1023 categories. In this case, there's a 0.9 chance of taking just one branch and a 0.1 chance of taking, on average, a little under 10 branches, so the average number of branches traversed in this tree is just under $1.9$ rather than the $10$ of the binary tree. The key insight is that the binary search tree minimises the \textit{maximum} number of branches that need to be taken, whereas we would like to minimise the \textit{average} number of branches. 

We can express this insight more formally by noting that the expectation value of the number of branches traversed in a single lookup is
\[
\mathbb{E}[L] = \sum_i \pi_i L_i
\]
where $L_i$ is the number of branches between the root and the $i^{th}$ category. If we now number each branch and let $d_{ji}=1$ if the $j^{th}$ branch lies between the $i^{th}$ category and the root of the tree, and $d_{ji}=0$ otherwise, then
\[
\mathbb{E}[L] = \sum_i \sum_j \pi_i d_{ji}
\]
now, letting $s_j = \sum_i \pi_i d_{ji}$
\[
\mathbb{E}[L] = \sum_j s_j
\]
That is, if each branch is associated with the sum of the probabilities of the categories that can be reached from that branch, then the expected number of branches for a lookup is the sum of these numbers over all branches.

 Huffman\cite{Huffman1952Method}, in the context of creating codewords for data compression, has described a method to construct a tree that is provably optimal with respect to this measure. In our context, the leaves of the tree represent categories from our distribution and internal ``sum-nodes'' are associated with the sum of the probabilities of their children. Given a sample from a uniform distribution, we can generate a sample from the categorical distribution in $\log(n)$ time using the algorithm in figure\ref{sampling}.
 
 \begin{figure}
 \begin{lstlisting}[frame=single, language=Python]
def sample(uniformSample) 
    node = root
    while(node is not a leaf)
        if(uniformSample < node.leftChild.value)
            node = node.leftChild
        else
            uniformSample = uniformSample - node.leftChild.value
            node = node.rightChild
    return(node)
 \end{lstlisting}
 \caption{Algorithm to generate a sample from a Huffman tree}
 \label{sampling}
\end{figure}
 
However, Huffman does not provide a method to efficiently add, delete or modify the probability of categories. More recently, the concept of adaptive Huffman coding has given rise to algorithms that allow modifications to the tree, for example, the Vitter\cite{vitter1989algorithm} and FGK\cite{knuth1985dynamic} algorithms. However, these algorithms assume the (un-normalised) probabilities can be represented as a integers, and only allows probabilities to be incremented or decremented. In our context we wish to allow arbitrary changes.

\section{Modifying parameters}

The algorithm to add a new category is given in figure\ref{addition}. Starting at the tree root, simply follow the least probable branch recursively until you reach a node whose probability is less than the new node's probability, then insert there. If the sum-nodes store their children in sorted order (i.e. with the less probable child to the right(left)), then one need only navigate down the right(left) edge of the tree.

 \begin{figure}
 \begin{lstlisting}[frame=single, language=Python]
def add(newNode) 
    node = root
    while(node is not a leaf and node.value > newNode.value)
    	node = node.lowestProbabilityChild
    newSumNode = Node(parent = node.parent, child1=node, child2=newNode)
    node.parent.swapChild(node, newSumNode)
    newSumNode.recalculateValueAndPropogateToRoot
 \end{lstlisting}
 \caption{Algorithm to add a new category to a tree}
 \label{addition}
\end{figure}

To delete a leaf-node, simply delete the sum-node directly above it and replace it with the deleted leaf-node's sibling.

Modification of the probability of a category can be achieved by deleting that category then re-adding it with the new probability.

Modification using these algorithms doesn't maintain the tree as a true Huffman tree, but we now show that, in practice, this is not a problem.

\subsection{Performance}

Table \ref{performance} shows the performance of the algorithm on a categorical distribution of approximately 100,000 categories and compares it to the theoretical optimum performance of a Huffman tree. These figures were calculated by initialising the tree with exactly 100,000 items with probabilities drawn at random, then performing a sequence of randomly generated addition, deletion and modification operations with uniform probability. When initialising the tree and adding/modifying categories, new probabilities need to be generated. These were drawn at random from a distribution. In order to test the algorithm under a range of circumstances, three different distributions were used to create the new values: A uniform distribution in the range $[0,1)$, an exponential distribution $P(x) = e^{-x}$ to simulate categorical distributions that are largely low probability with some higher probability peaks and a ``resonant'' distribution $P(x) = 0.99\delta(x-1) + 0.01\delta(x-1000)$ to simulate distributions that are predominantly low probability with a few, very sharp, resonant peaks.

After initialisation of the tree, a burn-in period of 250,000 operations was performed in order to reach equilibrium, then an additional 250,000 operations were performed during which, at every 500$^{th}$ operation, the optimal Huffman tree of the current categorical distribution was constructed and it's $\mathbb{E}[L]$ calculated along with the $\mathbb{E}[L]$ of the tree created by the algorithm described here.
 
Somewhat surprisingly for such a simple algorithm, the average time to do a single lookup, as measured by $\mathbb{E}[L]$, remains within a few percent of the theoretical optimum and remains so after multiple operations and over a range of probability distributions. This is because the algorithm to add categories promotes a roughly equal balance between the probabilities of each sum-node's children. When there are a large number of categories (i.e. when efficiency matters) there is necessarily a large number of categories with very small probability, so the sum-nodes closer to the root will be more finely balanced. Deletion tends to upset that balance, but any imbalance will be removed with subsequent additions. However, multiple deletions can, in the worst case, produce an unbalanced tree. We deal with that in the next section.

\subsection{The effect of tree rotations}

If a sum-node has a grand-child whose probability is greater than one of the sum-node's children, then a tree rotation towards the low-probability child would reduce $\mathbb{E}[L]$ (see for example\cite{cormen2009introduction} for a description of tree rotation). Upon addition and deletion, then, potential improvements could be made by checking for beneficial tree rotations between the modification point and the tree root. The result of doing this on the performance of the tree is shown in table\ref{performance}. Although performance is improved, the difference is small.

To test the effect of multiple deletions, we created a categorical distribution with 1,000,000 categories with probabilities chosen at random from a uniform distribution. Categories were then deleted at random until 1024 remained. Without rotation, $\mathbb{E}[L]$ was 1.0611 times the optimal Huffman value, whereas with rotation $\mathbb{E}[L]$ was 1.0211 times the optimal.

So, the expected improvement from doing tree rotations is small and unlikely to be worthwhile unless a large number of samples are to be taken between modifications or deletions.

\section{Code}

A Kotlin implementation of the algorithm described here, with and without tree rotations, along with the code used to generate the performance data, is available on Github at \href{https://github.com/danftang/MutableCategoricalDistribution}{https://github.com/danftang/MutableCategoricalDistribution}.

\begin{table}
 \caption{Average number of branches for one lookup of 100,000 categories using the algorithm described here (with and without tree rotation) compared to the theoretical optimum Huffman tree}
  \centering
  \begin{tabular}{lccccc}
    \toprule
    Distribution	&Optimal Length	& Measured  & Measured &	Ratio  &	Ratio \\
    &$\mathbb{E}_{opt}[L]$&$\mathbb{E}[L]$&$\mathbb{E}_{rot}[L]$ (with rotation)&$\frac{\mathbb{E}[L]}{\mathbb{E}_{opt}[L]}$&$\frac{\mathbb{E}_{rot}[L]}{\mathbb{E}_{opt}[L]}$\\
    \midrule
Uniform	&		16.3551&		16.4629	&	16.4265	&	1.0066	&	1.0044\\
Exponential&		16.0282	&	16.2049	&	16.1377	&	1.0110	&	1.0068\\
Resonance	&	10.9817	&	11.4774	&	11.3389	&	1.0451	&	1.0325\\
    \bottomrule
  \end{tabular}
  \label{performance}
\end{table}

\bibliographystyle{unsrt}  
\bibliography{references}

\end{document}